# The mechanism of tornadogenesis from the perspective of vortex tubes

Peng Yue[1], Y. Charles Li[2], Jiamin Dang[1], Leigh Orf[3], Grace Yan[1]

1. Department of Civil, Architectural and Environmental Engineering, Missouri University of Science and Technology, Rolla, MO 65401, USA
2. Department of Mathematics, University of Missouri, Columbia, MO 65211, USA
3. Space Science and Engineering Center, University of Wisconsin-Madison, Madison, WI 53715, USA

## 1. Introduction

Most high-intensity tornadoes are associated with supercells, even though other weather systems such as squall lines can generate tornadoes too. Roughly 20%-30% of supercells produce tornadoes. The main components of a supercell are shown in Figure 1. Supercells are generated from horizontal veering vortices uplifted by warm updrafts (See Figure 2). There are two theories with regard to near ground dynamics of supercells. One believes that the supercells end at the wall clouds level, below which there is no vorticity. The other believes that the supercells are blown by the Rear Flank Downdrafts (RFD) to the ground. We believe in the latter theory. From the vortex tube perspective, supercell vortex tubes cannot stop midair, and they have to connect to the ground (See Figure 2). By the Kelvin-Helmholtz Theorems [1], the vortex flux is constant along all horizontal sections of the vortex tube. When the vortex tube approaches the ground, its horizontal sectional areas may get much bigger, thus the vorticity is much smaller. Since the tornado flow is turbulent, the vortex tubes are also turbulent, thus the horizontal sections in general are fractal. The sizes of the areas of such fractal sections sometimes can be counter intuitive, i.e. the fractal sections may look big, but have very small areas. Even with fractal horizontal sections of the vortex tubes, when the outer pressure of the vortex tube is larger, the areas of the horizontal sections will be squeezed to decrease.

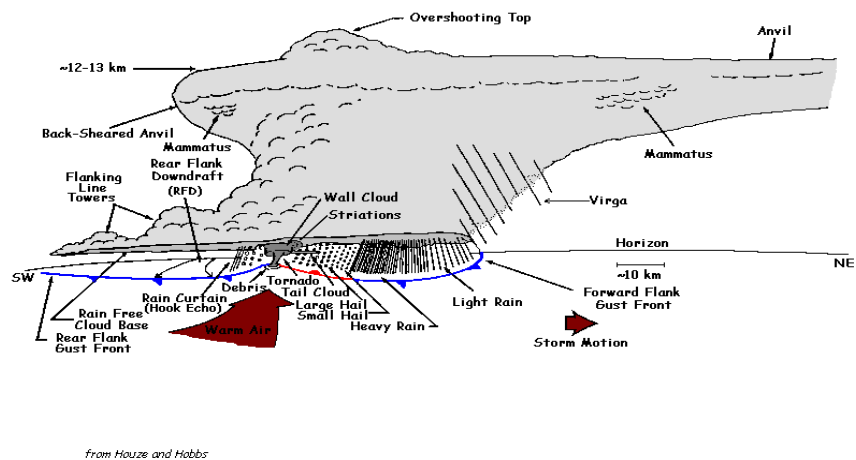

Fig. 1 Supercell storm's main components [2].

Tornadogenesis is currently an active area of research. There are two vague theories on tornadogenesis. One involves a vague theory of some sort of vortex stretching [3]. The other proposes that tornadoes are generated by new horizontal vortices created near ground, uplifted by the updraft in a similar manner with supercells, and attached to the supercells' wall clouds [3]. In this paper, we propose a new theory: We start with supercells blown by the Rear Flank Downdrafts (RFD) to the ground. The RFD cool front and the updraft warm front (which is the same with the cool front of the Forward Flank Downdraft (FFD)) form a tilted wedge shape, see Figure 2b.The pressure in the warm updraft region is high, while the pressure in the cool RFD region is low. The lowest pressure line in the RFD region near the intersection tip of the RFD cool front and the updraft warm front will form the center of the possible tornado. The pressure difference between the lowest pressure line and its surroundings causes the vortex tubes centered around the lowest pressure line being squeezed, and their horizontal cross sections to shrink. By the Kelvin-Helmholtz Theorems on vortex tubes, the vortex flux is constant under the fluid motion, thus the vorticity on the squeezed sections increases. When the pressure difference between the lowest pressure line and its surroundings is large enough, the increase of vorticity can reach the tornado level, and thus a tornado is born. When the pressure difference increases, the tornado strength increases. When the pressure difference decreases, the tornado strength decreases. The decay of tornadoes is caused by the decreasing pressure difference. This is our theory of the entire tornado lifespan.

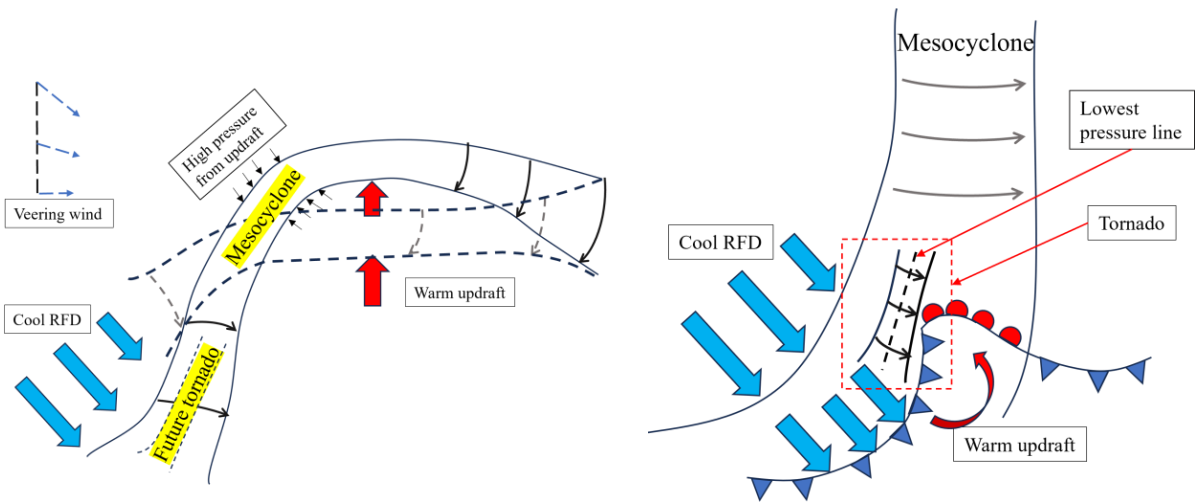

a) Due to veering wind shear, air current spinning is formed horizontally with vortex tubes veering along the storm direction, the updraft continuously pushes the vortex tube upward, and the rear flank downdraft (RFD) pushes one end of the vortex tube downward toward the ground.

b) A not too cold RFD can generate a large enough pressure difference between the lowest pressure line and its surrounding, which will squeeze the vortex tube area to be smaller enough, leading to higher enough vorticity resulting in a visible tornado.

Fig. 2 New theory of tornadogenesis based on vortex tubes.

With our theory on tornadogenesis, we can offer a resolution to the conjecture on the RFD effects on tornadoes posed in [4]: When the RFD is cool but not cold, the pressure difference between the lowest pressure line and its surroundings is very large, and it creates a strong tornado. When the RFD gets colder, the pressure difference decreases, and the tornado is weakened. When the RFD gets very cold, the pressure difference is not large enough to support a tornado.

## 2. Verification of our theory

If any closed curve in the fluid flow field is picked, all the vortex lines connecting to the closed curve form a vortex tube. If the vortex lines are transversal to the closed curve, then the vortex tube bears the regular tube shape in the neighborhood of the closed curve. Since the tornado fluid flow is turbulent, globally the vortex tube is very irregular. Nevertheless, for any tornado turbulent flow, the vortex tube can still be calculated and plotted.

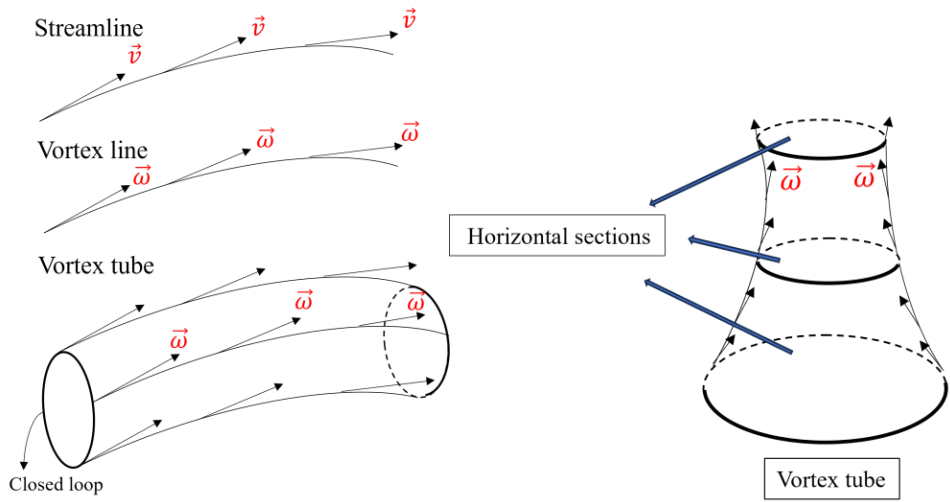

Fig. 3 Schematic diagram of streamline, vortex line and vortex tube.

2.1 Vortex tubes

Choosing a horizontal circle around a lowest pressure point near the wall clouds, we can generate a downward vortex tube which bears a brush form. Clearly the brush form is blown by RFD down to the ground. As time goes on, the brush form is further blown by RFD to rotate counter clockwise near the ground (See Figure 4).

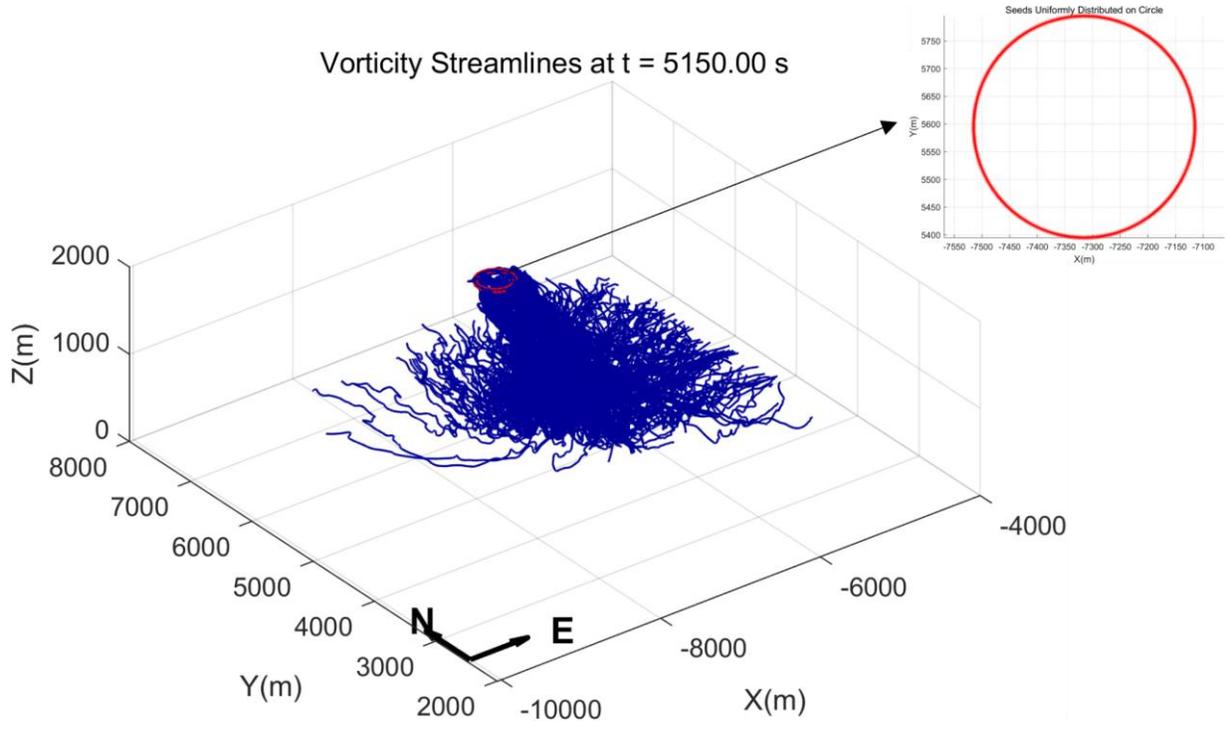
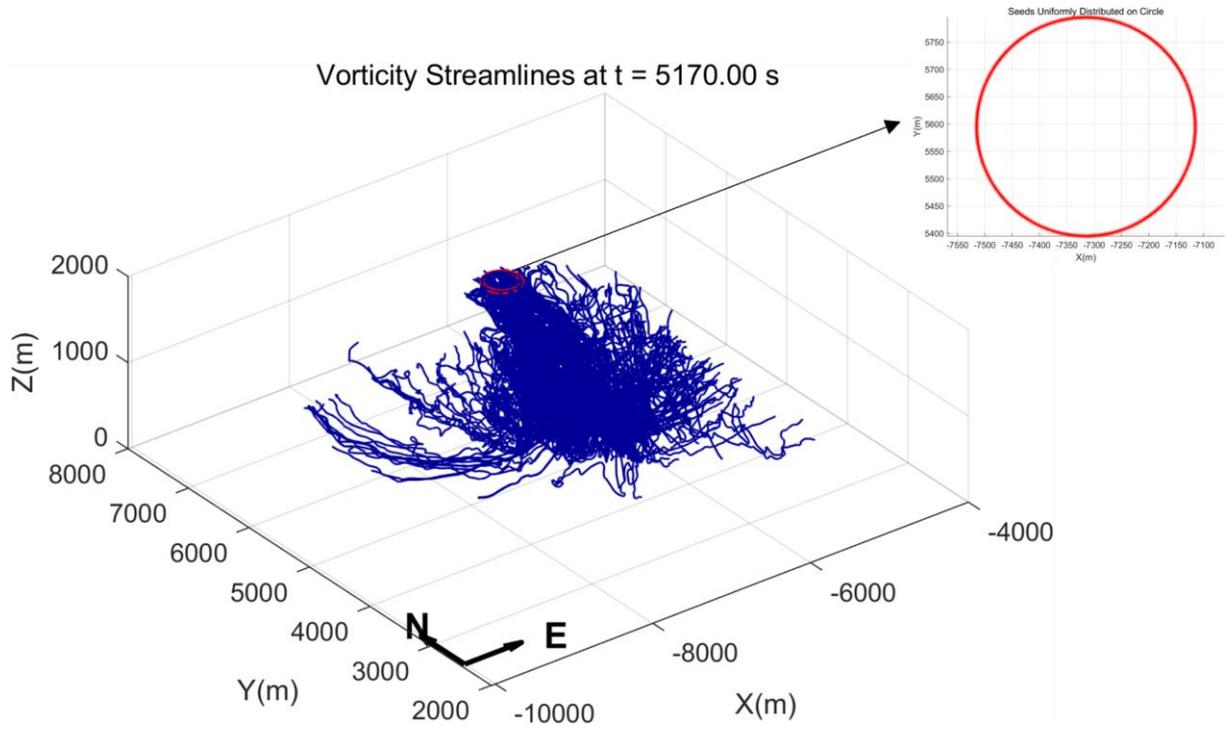

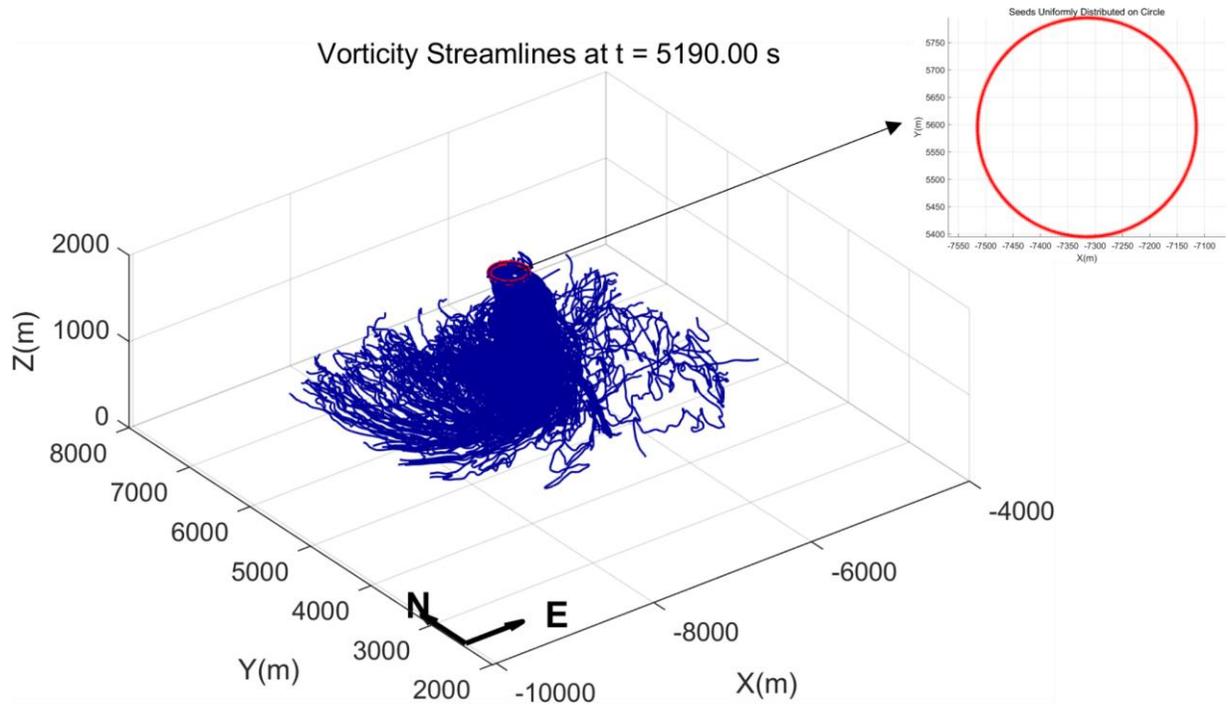

Fig. 4 The vortex tube before the formation of a tornado.

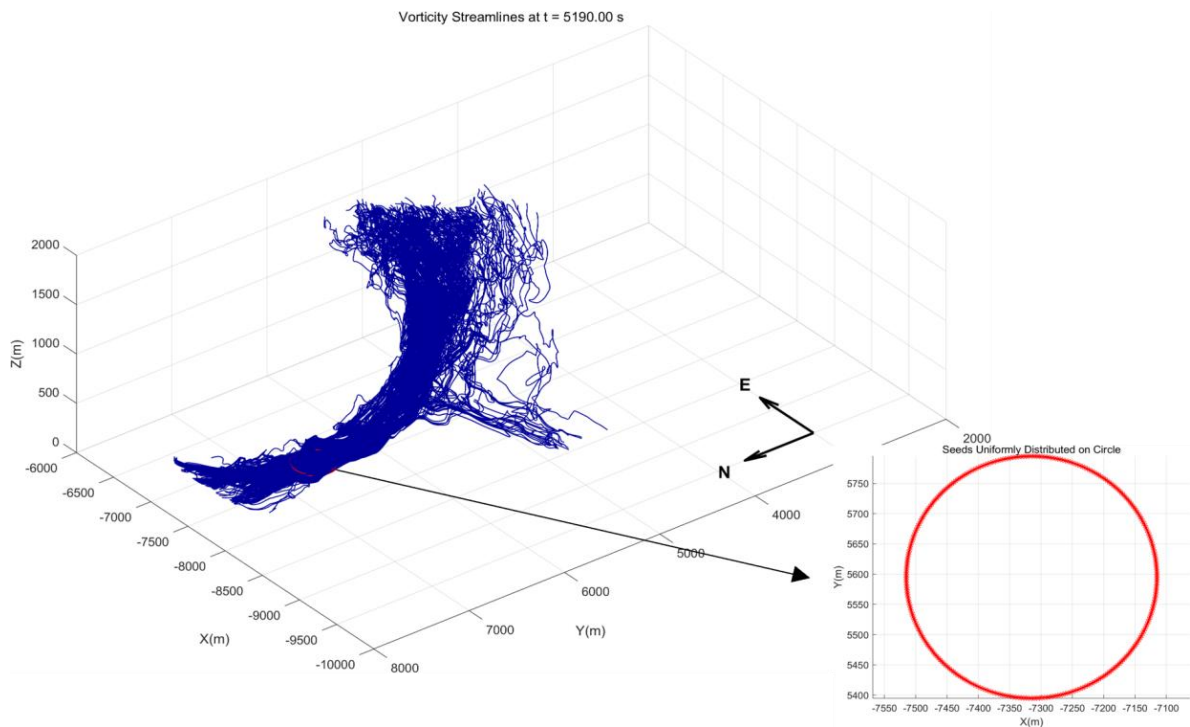

Fig. 5 The vortex tube connecting to a horizontal circle near the ground, which still bears a brush shape.

This serves as a verification that the supercells are blown downward to the ground by RFD. The horizontal sections of the turbulent vortex tube are sort of fractal. The areas of such fractal sections can sometimes be counterintuitive: Those that look large may be actually small, while those that look small may be relatively large. The areas of the horizontal sections of the downward vortex tube are increasing downward. Thus the vorticity is decreasing downward. We can choose a horizontal circle near the ground and generate an upward vortex tube which still bears a brush form since the flow is turbulent (See Figure 5). The areas of the horizontal sections of the upward vortex tube are also increasing downward, even though they may look counterintuitive. Thus the vorticity is decreasing downward.

We can also generate a solid vortex tube connecting to the horizontal disc with the horizontal circle as its boundary. One can see the solid vortex tube looks very similar to the vortex tube connecting to the boundary circle. This further implies the fractal nature of the horizontal sections of the vortex tube (See Figure 6).

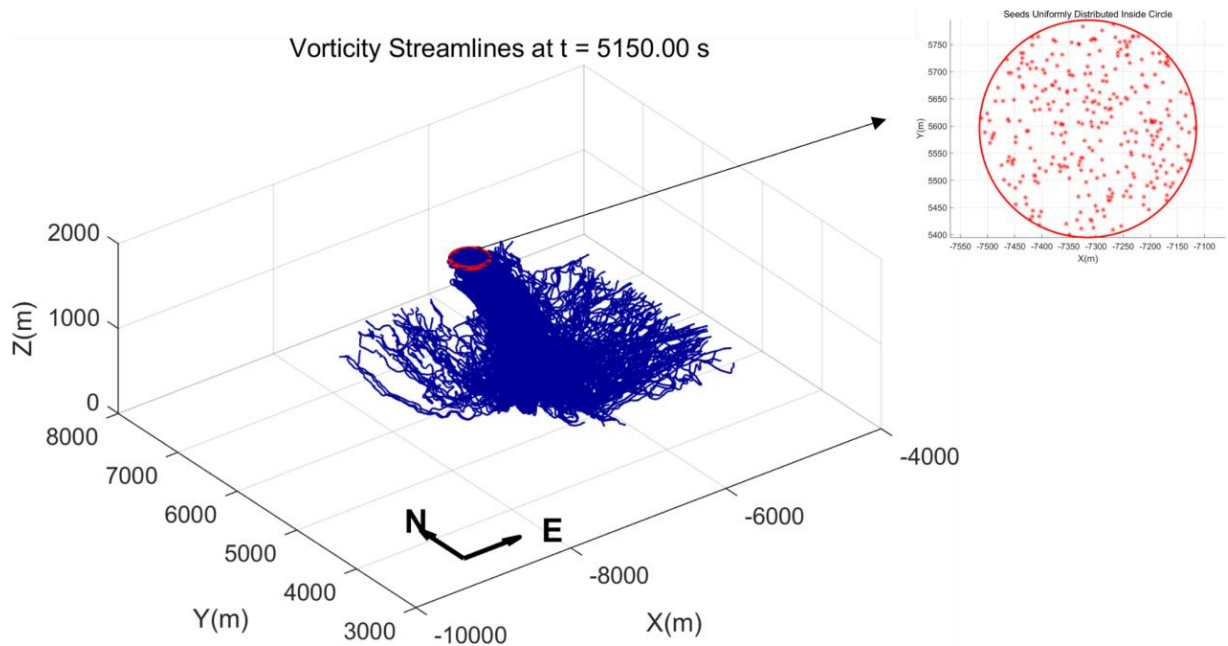

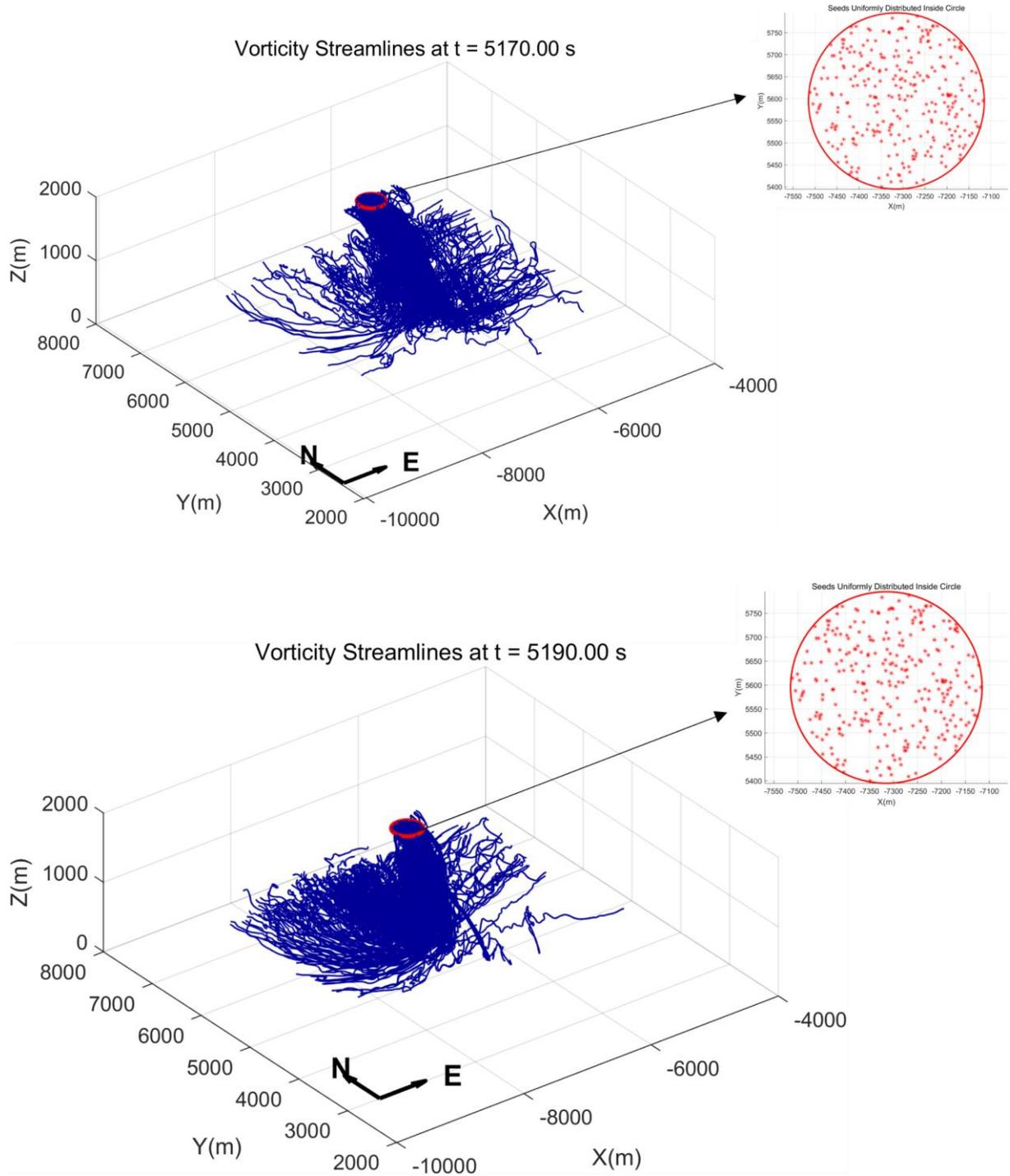

Fig. 6 The solid vortex tube before the formation of a tornado.

The vortex tube appears more like a conventional tube only after the tornado has fully formed and a smaller horizontal circle is selected (See Figure 7).

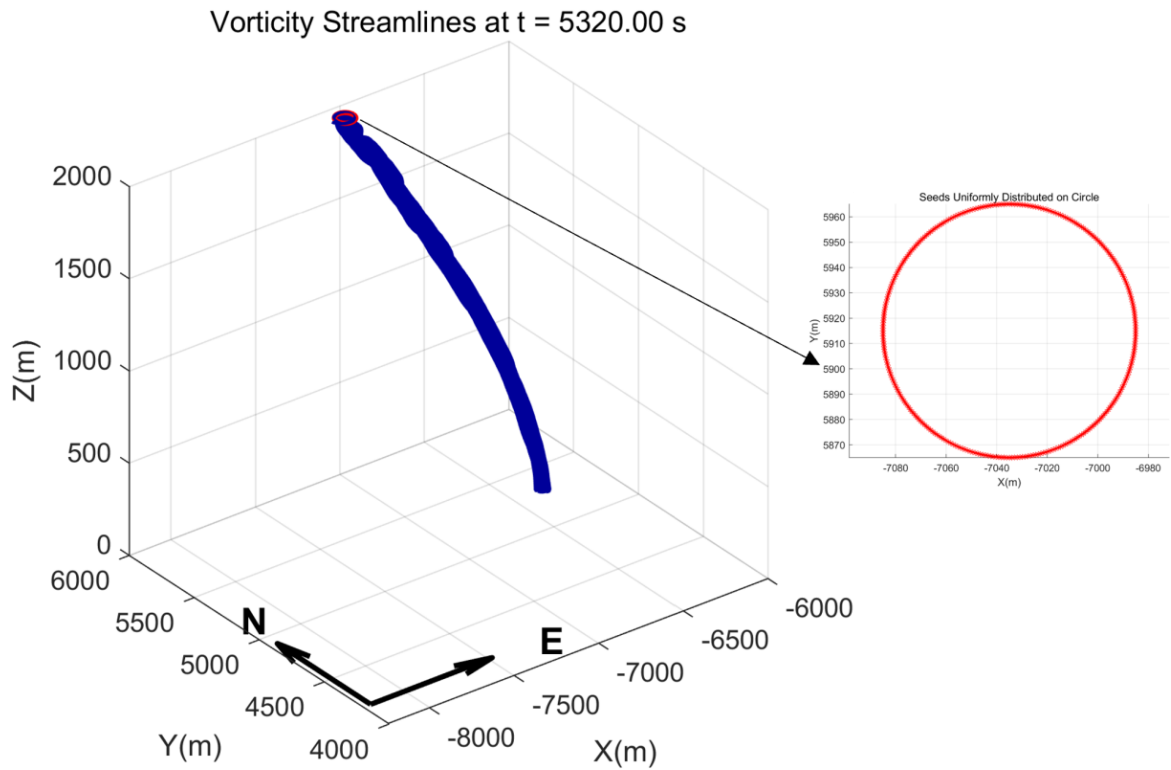

Fig. 7 The vortex tube formed after the formation of a tornado.

2.2 Pressure

To verify our theory on tornadogenesis, we plot the pressure field around the vortex tube. The lowest pressure line is along the center of the vortex tube. The pressure outside the vortex tube is greater than that inside the vortex tube. Before the appearance of the tornado, the pressure difference between the outside and the inside of the vortex is relatively small. When the pressure difference is large enough, a tornado appears. As the pressure difference increases, the tornado strengthens. As the pressure difference decreases, the tornado weakens. When the pressure difference is small enough, the tornado disappears (See Figure 8).

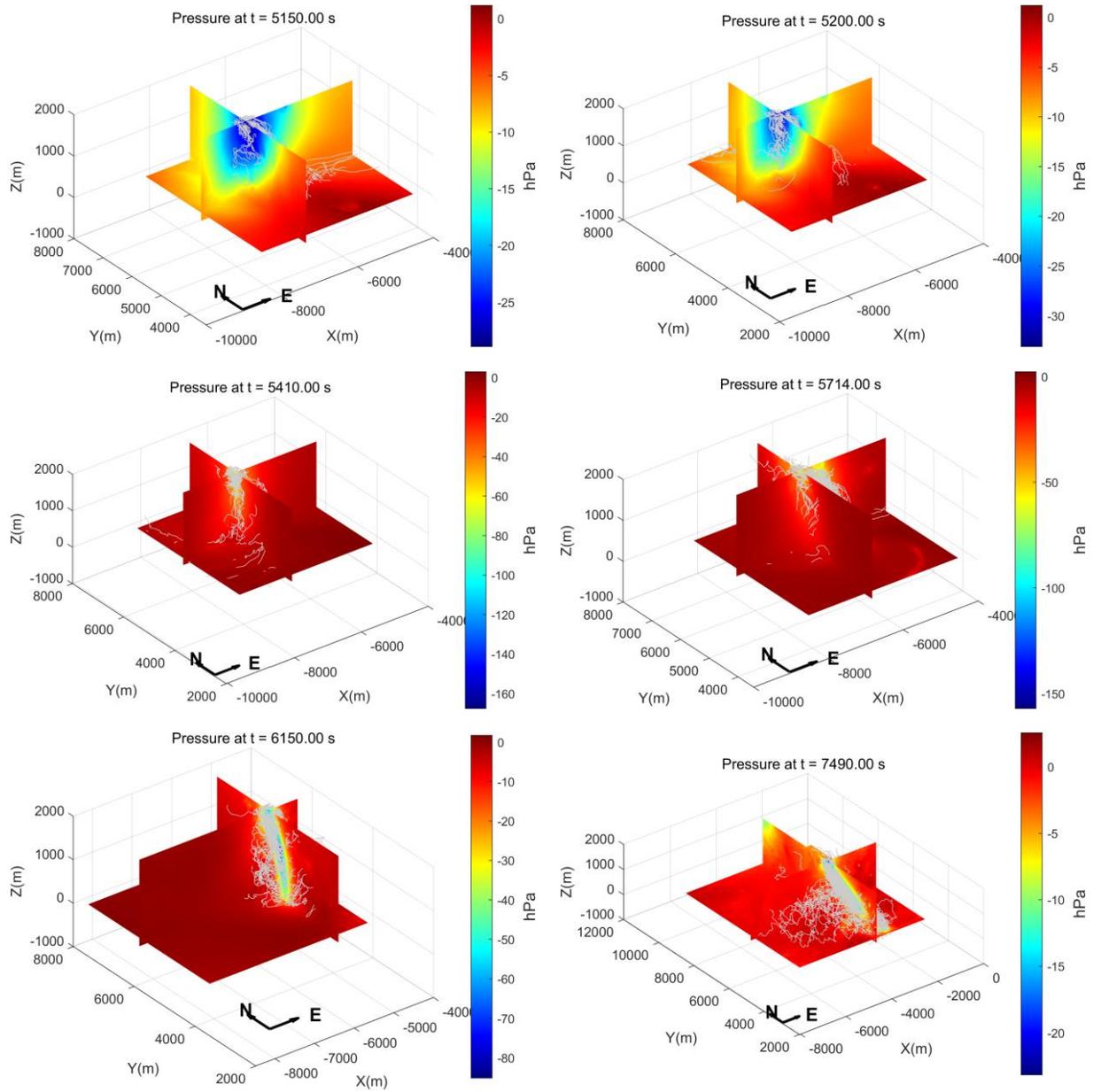

Fig. 8 The pressure distribution throughout the entire life cycle of the tornado.

## 2.3 Temperature

Since the gas density is not constant, pressure and temperature are not proportional. Figures 9 show the corresponding temperature field distributions.

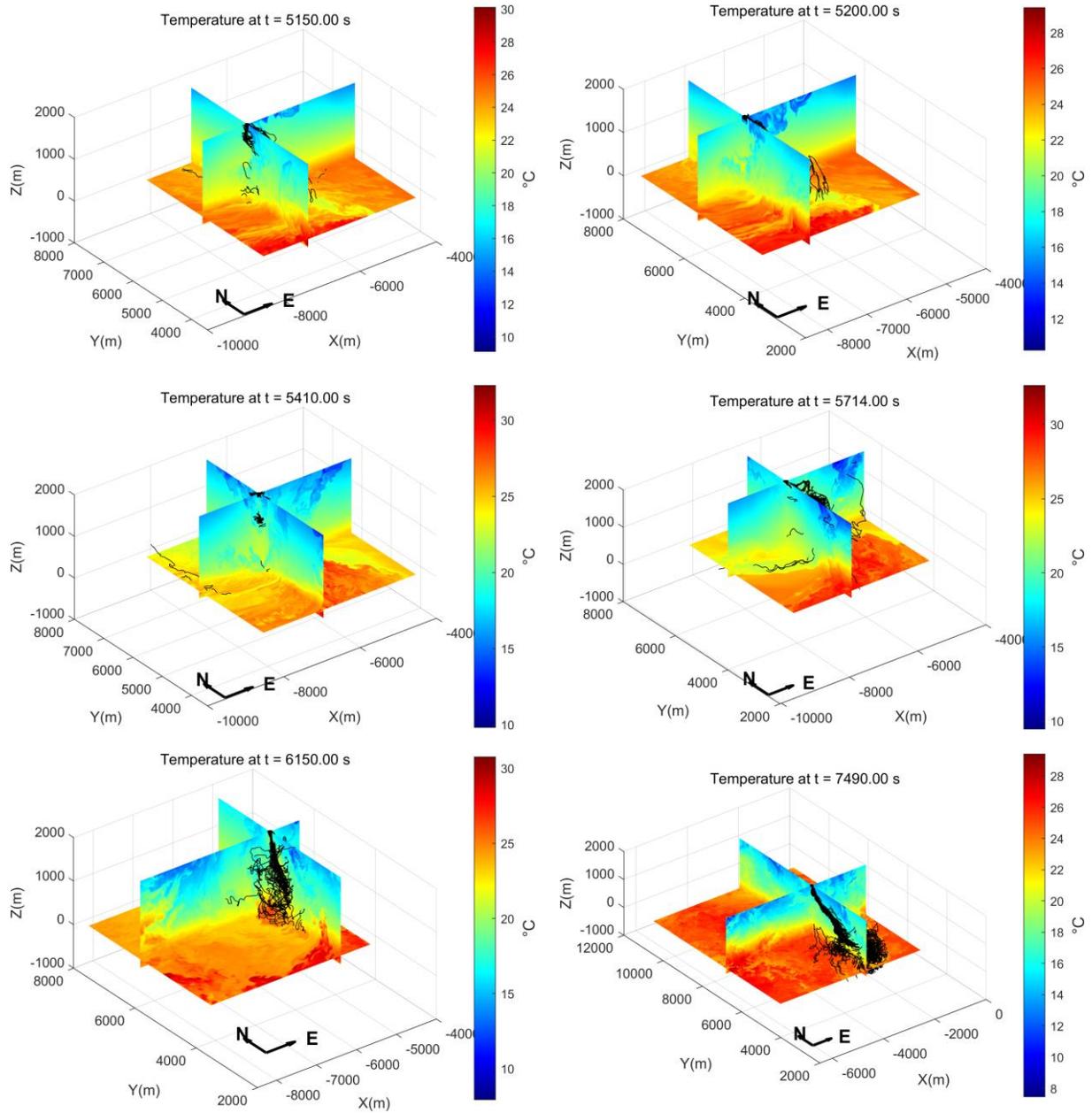

Fig. 9 The temperature distribution throughout the entire life cycle of the tornado.

## 3. Conclusion

This is a preliminary version of our theory on tornadogenesis and its life span. The comprehensive version of this work is underway. We believe that the same mechanism also applies to the formation of the supercells, that is, the supercells are formed via pressure squeezing from the high pressure warm updrafts surrounding the supercell vortices. See Figure 2(a).